# 'Boson peak' and high-frequency excitations in glassy crystals

P. Lunkenheimer and A. Loidl

*Experimentalphysik V, Institut für Physik, Universität Augsburg, 86135 Augsburg, Germany*

**The high-frequency excitations in glasses and supercooled liquids belong to the great mysteries of the physics of condensed matter. While the fast process, located at GHz-THz, can be interpreted as encaged molecular motion, the occurrence of the boson peak (BP), dominating at THz frequencies, still is controversially debated. Two scenarios have evolved during recent years: the BP is truly collective in nature[1,2,3,4] or, alternatively, it results from localized modes[5,6]. To help solving this controversy, we have investigated the high-frequency dielectric loss of plastic crystals (PCs), having long-range translational but no orientational order. Having in mind the well-defined phonon modes in PCs compared to canonical glasses, and the infrared silence of acoustic phonon modes, millimeter-wave and far-infrared spectroscopy should provide valuable hints on the true origin of the BP. Here we show that PCs exhibit a BP-like feature with markedly different spectral shape compared to canonical glasses, evidencing that mixing of collective phonon with local relaxational modes causes the BP. In addition, we find clear evidence for a fast process also in PCs.**

We have investigated the frequency-dependent dielectric loss $e''$ of the prototypical plastic crystals ortho-carborane (OCA) and 1-cyanoadamantane (CNA). The nearly spherically shaped OCA molecule, $B_{10}C_2H_{12}$, forms a rigid icosahedron. In CNA ($C_{10}H_{15}CN$) the carbon atoms make up a rigid cage with one cyano side-group. Measurements in an exceptionally broad frequency range were performed by a combination of different experimental techniques, employing coaxial reflection and transmission techniques, submillimeter wave spectroscopy, and Fourier transform infrared spectroscopy (for details, see ref. 7).

Figures 1 and 2 show the loss spectra of OCA and CNA for $\nu > 100$ MHz up to infrared frequencies. For the highest temperature, below 10 GHz the so-called α-peak shows up, which marks the characteristic time scale of the reorientational dynamics of the dipolar molecules. As known from previous work[8,9,10,11,12,13], with decreasing temperature the α-peak shifts towards lower frequencies, mirroring the glass-like slowing down of the molecular dynamics. It is well known that the α-relaxational dynamics in plastic crystals, reflecting reorientational molecular motions, exhibits completely analogous behavior to that in canonical glass-formers, where it is coupled to translational motions, a fact that led to the establishment of plastic crystals as model systems for canonical glass-formers. As revealed by Figs. 1 and 2, this analogy seems to prevail even for the high-frequency dynamics: Around 100 GHz, very similar to canonical glass-formers[14,15] both plastic crystals exhibit an $\varepsilon''(\omega)$-minimum with clear indications for an additional fast process. Namely at the lower temperatures, for both materials, a simple superposition of two power laws, determined by the high-frequency wing of the α-peak and the increase towards the peak in the THz region (dashed lines in Figs. 1 and 2), cannot describe the experimental data satisfactorily. These results give a strong hint, that also in plastic crystals there is an additional fast process contributing to the loss in this region. In canonical glass formers, the fast process often is ascribed to the motion of the molecules in the cage formed by the adjacent molecules, as described in detail by the mode coupling theory (MCT) of the glass transition[16]. Some indications for the applicability of the MCT to plastic crystals were obtained recently[17], but of course it remains to be clarified what could be the analogue to the cage effect in plastic crystals. To account for the excess intensity in the minimum region without invoking a specific model, the data were parameterized by adding a constant to the two power laws (solid lines). It should be noted that such a constant-loss contribution was also employed to describe the high-frequency response of canonical glass-formers[14,18].

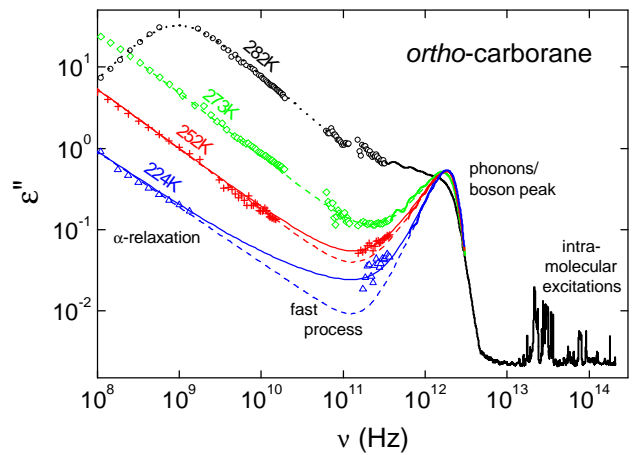

**Figure 1** Dielectric loss spectra of plastic-crystalline OCA for various temperatures. To maintain readability, at $\nu > 400$ GHz the data are shown as solid lines and at $\nu > 3$ THz only data for 282 K are shown. The dashed lines are calculated by a superposition of two power laws; the solid lines comprise an additional constant loss contribution. The dotted line is a guide to the eye.

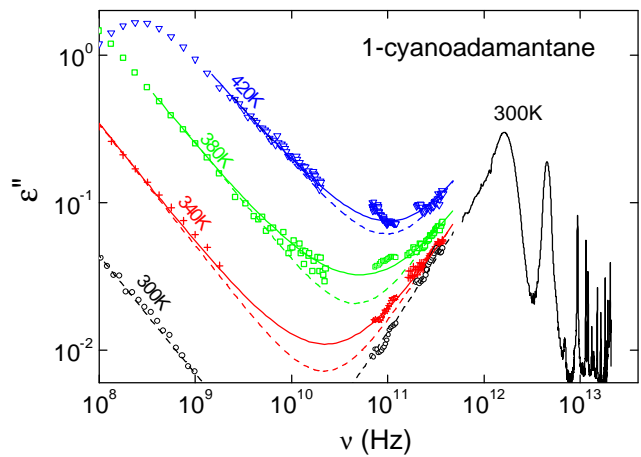

**Figure 2** Dielectric loss spectra of plastic-crystalline CNA for various temperatures. The meaning of the lines in the minimum region is the same as in Fig. 1.

Near 2 THz, an only weakly temperature-dependent peak shows up in both plastic crystals (Figs. 1 and 2). This peak is rather broad, with a relatively moderately increasing low-frequency wing, and thus clearly is not due to a resonance-like excitation, as, e.g., a phonon mode or an intramolecular vibration. Thus this peak seems to represent the analogue to the BP, observed in canonical glass formers.[15,19,20] At higher frequencies, further, much sharper peaks show up. These resonance-like modes can be ascribed to intramolecular excitations[21,22], e.g. in CNA the most prominent of these modes at about 4.5 THz was ascribed to the bending motion of the C-C≡N group[22].

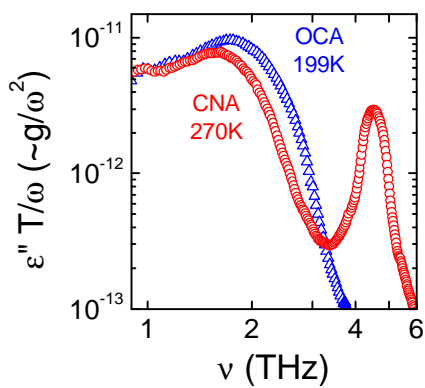

**Figure 3** Spectra of $\varepsilon''T/\omega$ in the THz region. The plotted quantity is approximately proportional to the DOS, divided by the squared frequency, $g(\omega)/\omega^2$. Shown are the spectra of plastic-crystalline OCA and CNA at the lowest temperatures investigated.

In literature, the BP is often defined as a peak in the density of states (DOS) divided by the squared frequency, $g(\omega)/\omega^2$, implying an excess over the Debye vibrational DOS. Figure 3 demonstrates for the lowest temperatures investigated, where contributions from α-relaxation and fast process can be neglected, that the detected boson-peak-like feature in PCs also corresponds to a peak in the quantity $\varepsilon''T/\omega$, which is a good approximation of $g(\omega)/\omega^2$. In most reports on the BP of canonical glass formers, $g(\omega)$ is extracted from neutron or light scattering experiments, which mainly couple to translational degrees of freedom. However, it was demonstrated[15] that FIR spectroscopy, despite coupling predominantly to reorientational motions, leads to a very similar spectral shape of the BP.

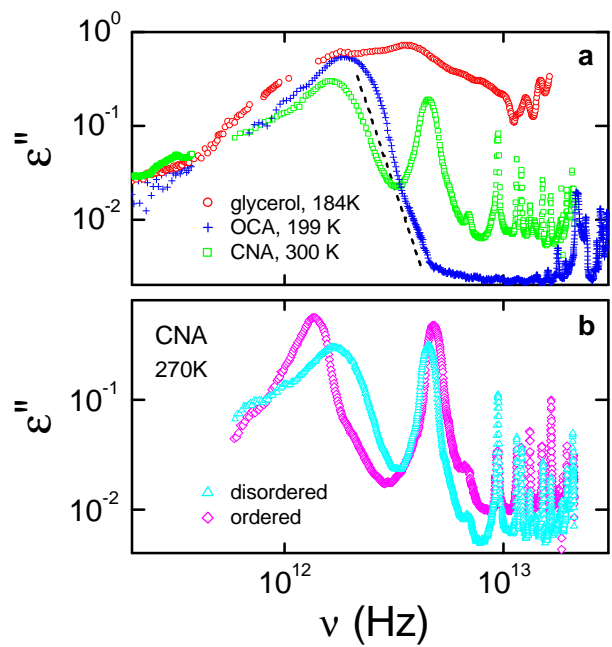

**Figure 4** Comparison of the THz loss-peaks in completely disordered, orientationally disordered, and completely ordered materials. **a,** Dielectric loss spectra of the plastic crystals OCA and CNA and the canonical glass former glycerol. The dashed line represents a power law, $\varepsilon'' \sim \nu^{-7}$. **b,** Dielectric loss spectra of orientationally disordered and ordered CNA at 270 K.

In Fig. 4a the BP in the canonical glass former glycerol[23] is compared to the corresponding peaks in OCA and CNA. Again low temperatures were chosen to exclude any influence from processes at lower frequencies. The low-frequency wing of the THz-peak for all three materials looks quite similar, increasing with an approximate power law with an exponent of 1 - 1.5. However, following an approximate power law with exponent of -7 (dashed line in Fig. 4a), the high-frequency wing for the plastic crystals exhibits a much steeper decrease than in glycerol, where $\varepsilon''(\omega)$ decreases with similar slope as on the left wing, before the intramolecular modes come into play above about 10 THz. This strong dissimilarity of the spectral shape in canonical glasses and plastic crystals reveals that phononic, i.e. collective modes must play an important role in the formation of the BP: In plastic crystals a BP still is present, due to their disordered nature, but the well-defined phonon modes in these materials lead to a different spectral shape. But what causes the observed strong *asymmetry* of the peak in plastic crystals? A possible explanation emerges invoking ideas that the BP may simply reflect the complete phonon DOS, becoming IR active via the strong hybridization of collective phonon modes with local relaxational excitations[24,25,26]. OCA can be viewed analogous to a monoatomic crystal (hard spheres of $B_{10}C_2H_{12}$, linked by relatively weak van-der-Waals bonds), with a purely acoustic Debye phonon DOS. Then the high-frequency cut-off of the Debye DOS is the cause of the observed strong decrease at the high-frequency wing of the THz-peak. In plastic-crystalline CNA, due to the less rigid molecular structure, the intramolecular modes arise at lower frequencies, a side-band excitation appearing close to 4.5 THz; however, the very similar shape and amplitude of the THz peak (Fig. 4a) indicates a similar underlying physics. In contrast to the plastic crystals, in canonical glass-formers all phonon modes, considerably smeared out due to the lack of long-range translational order, and low-frequency intramolecular modes mix up to form a peak with a much softer high-frequency wing. Also optical phonon modes may play a role here, the observed multi-peak structure (Fig. 4a) resulting from the complete vibrational DOS. Nevertheless, it is reasonable that the low-frequency part is dominated by the acoustic DOS only and thus is similar to that of the plastic crystals.

The concept, that the BP results from a strong mixing of localized and extended modes can further be checked in CNA. CNA can be prepared in glassy and in ordered form, when annealed at low temperatures for several hours. Figure 4b shows the results for completely ordered and orientationally disordered CNA, both at 270 K. Interestingly, in the ordered state again a peak shows up close to THz, but now located at somewhat lower frequencies. This peak, having a significantly smaller width than in the disordered state and being well describable by a Lorentz function, shows the typical signature of a resonant excitation. In the completely ordered state, where any contributions from the purely acoustic phonon DOS can be excluded, the only feasible explanation for this feature is a librational excitation. In the plastic phase these localized modes become strongly damped and the excitation spectrum becomes relaxational in nature. The relaxation slows down with decreasing temperature, primarily constituting the α-relaxation spectrum in Figs. 1 and 2. But their coupling to and mixing with the acoustic modes via local strain fields constitutes the BP.

To summarize, from dielectric spectroscopy on PCs we were able to clarify the origin of the BP in disordered matter: At first, the mere fact that we detect a BP with dielectric spectroscopy, which is only sensitive to molecular reorientations, proves that it cannot be due to acoustic phonon modes only. Instead, either local relaxational or collective optic modes must play a role and if acoustic modes are involved at all, this can only happen via hybridization with reorientational modes. The detected strong asymmetry of the BP in PCs, which is in marked contrast to the spectral shape found in canonical glasses, in accord with arguments in favor of a mainly acoustic character of phonon modes in the investigated PCs, strongly indicates that indeed acoustic modes dominate the BP in CNA and OCA. Thus the fact that the BP is infrared active lets us conclude that the appearance of a BP in CNA and OCA results from a strong mixing of relaxational localized and phonon-like extended modes. This conclusion is further corroborated by our results on CNA in its completely ordered phase. It is reasonable that similar effects are also responsible for the BP of structurally disordered materials.

**Acknowledgements**

We thank R. Brand and R. Wehn for performing part of the measurements.